\documentclass[]{elsarticle}

\makeatletter
\def\ps@pprintTitle{%
  \let\@oddhead\@empty
  \let\@evenhead\@empty
  \let\@oddfoot\@empty
  \let\@evenfoot\@oddfoot
}
\makeatother

\usepackage{listings}
\usepackage{xcolor}
\usepackage{multirow}
\usepackage{booktabs}
\usepackage[hyphens]{url}
\usepackage[switch]{lineno}

\definecolor{mauve}{rgb}{0.58,0,0.82}
\begin{document}

\begin{frontmatter}

\title{Automated Generation of High-Performance Computational Fluid Dynamics Codes}

\author[bsc]{Sandra Maci\`a\corref{cor1}}
\ead{sandra.macia@bsc.es}
\cortext[cor1]{Corresponding author}

\author[bsc]{Pedro J. Mart\'inez-Ferrer}
\ead{pedro.martinez-ferrer@bsc.es}

\author[bsc]{Eduard Ayguad\'e}
\ead{eduard.ayguade@bsc.es}

  \author[bsc]{Vicen\c{c} Beltran}
\ead{vicenc.beltran@bsc.es}

\address[bsc]{Barcelona Supercomputing Center (BSC-CNS), Barcelona, Spain}

\begin{abstract}
Domain-Specific Languages (DSLs) improve programmers productivity by decoupling problem descriptions from algorithmic implementations. However, DSLs for High-Performance Computing (HPC) have two additional critical requirements: performance and scalability.
This paper presents the automated process of generating, from abstract mathematical specifications of Computational Fluid Dynamics (CFD) problems, optimised parallel codes that perform and scale as manually optimised ones.
We consciously combine within Saiph, a DSL for solving CFD problems, low-level optimisations and parallelisation strategies, enabling high-performance single-core executions which effectively scale to multi-core and distributed environments.
Our results demonstrate how high-level DSLs can offer competitive performance by transparently leveraging state-of-the-art HPC techniques.
\end{abstract}


\end{frontmatter}

\section{Introduction}
Scientific applications face the challenge of efficiently exploiting increasingly complex parallel and distributed systems.
Extracting high performance requires deep expertise in parallel programming models, libraries and algorithms, and in-depth knowledge of the target architecture.
Hand-tuned codes are built under this assumed knowledge, and therefore, able to provide both low-level optimised and efficient parallel implementations.
Data structures, optimisations and parallelisation strategies are intertwined with the application code and exposed to the compilers.
Nevertheless, developing such codes is a time-consuming, tedious and hardly reusable task.
In this scenario, reaching high performance appears detrimental to productivity and portability and unreasonable to expect from scientists.
Domain-Specific Languages (DSLs) have arisen as a separation of concerns approach through high-level abstraction layers to overcome such difficulties.
On the one hand, productivity and portability can be reached by abstracting the application layer from the final parallel low-level code.
On the other hand, since DSLs are restricted to specific problem domains, they are built under assumptions, embodying domain knowledge that enables the automated application of suitable computing methodologies. Hence, DSL can transparently tackle performance at the set of algorithmic patterns they implement.
However, this dissociation might blind the compilers and prevent using state-of-the-art HPC techniques such as vectorisation and tiling or advanced parallelisation strategies.
Thus, pressure has moved to DSLs researchers expecting from their framework a high level of generality and abstraction while delivering high-performance on par with hand-tuned codes.

This paper aims to enhance a DSL framework to automatically generate code optimised to exploit an HPC cluster from high-level specifications.
Saiph is a DSL that targets the resolution of Computational Fluid Dynamics (CFD) problems through a high-level syntax and a generic numerical library implementing explicit and Finite Differences Methods (FDM).
We use Saiph~\cite{saiph:rwdsl, saiph:pasc} as a DSL platform and consequently focus on the CFD domain.
However, the modular design of the tool enables the present methodologies to be generalised for other problem domains according to their computational needs.
We profit from the layered design and adapt the build process for a new code generation, ensuring the extraction and propagation of information from the input code to the final binary.
From Saiph applications, we automate the combination of state-of-the-art and advanced HPC techniques by generating an intermediate annotated code exposing detailed information to the compiler.
For that, we study such techniques and determine (i) the information needed at the intermediate code to enable it, (ii) how to extract and generate such information from a high-level specification and (iii) how to combine it with the rest of the techniques.

In particular, the combination of HPC techniques is a non-trivial task.
Simultaneously vectorising, blocking, parallelising and distributing a loop requires a harmonious combination ensuring aligned data access, good cache locality and well-balanced domain partition at the same time.
Our approach ensures the automated, generic and effective combination of such techniques by coupling the efficient exploitation of three levels of HPC resources: single-core, multi-core and cluster, in a bottom-up manner:

\begin{description}
\item[Single-core performance] We target single-core performance by exploiting parallelism and memory hierarchy through code-vectorisation and cache locality enhancement using data-blocking mechanisms.
To that end, we explore suitable data layout, traversal and alignment, compiler hints and compile-time evaluations.
\item[Multi-core performance] We inquire how to adapt the data-blocking for subsequent appropriate shared-memory usage at the node level when exploiting multi-core resources.
We target different shared-memory models such as fork-join and tasking through OpenMP and OmpSs-2~\cite{pm:ompSs} programming models. Hence, we address multi-core parallelism for load imbalance or data locality improvements.
To the best of our knowledge, this work presents an unprecedented approach for the automatic generation of taskified codes within a DSL framework.
\item[Distributed performance] On top of it, we research how to extend distributed executions based on MPI and scalable domain decomposition to preserve the underlying optimisations. Moreover, combining distributed strategies with multi-core parallel versions produces hybrid configurations with different interoperability characteristics. We study such hybrid configurations using MPI and TAMPI~\cite{tampi} combined with fork-join and task models.
\end{description}

The paper makes the following contributions:
\begin{enumerate}
  \item A DSL build process and code generation ensuring automated code annotations and optimisations.
  \item An effective automated combination of state-of-the-art HPC techniques and parallelisation strategies at single/multi-core and cluster levels.

\end{enumerate}

\section{Background}

\subsection{Computational Fluid Dynamics}
CFD is a physics domain to solve problems related to fluid flows numerically.
The governing equations modelling density-based CFD problems are the Navier-Stokes equations and the equation of state.
The former is a set of space-time dependent Partial Differential Equations (PDEs) describing the motion of a fluid; the latter corresponds to a non-time derivative equation relating the fluid's state variables.
Depending on the equations' unknowns, we incorporate other thermodynamic relations between state variables to close the system.
To unambiguously define a CFD problem, we state the equations' system, the space-time modelling scope and the initial and boundary fluid conditions (ICs, BCs). We use space-time discretisation methods for CFD system resolution.

\subsubsection{Explicit and Finite Difference Method (FDM)} \label{back}
Explicit FDM defines a closed set of computation patterns involving finite difference discretisations and explicit time solvers to solve a wide range of CFD problems.
FDM discretises a spatial domain mapping continuous field information into a Cartesian grid of points; points store field values according to spatial coordinates.
Explicit methods discretise the temporal dimension through a time-stepping loop at which each iteration defines the state of a system using previous time-step states.
In such methods, PDEs are approximated by algebraic equations at each spatial coordinate at each time-step.
Computationally, a time-step loop encloses a spatial traversal at which linear algebra and stencil computations occur.

\subsubsection{Challenges of HPC explicit FDM} \label{backChallenges}
Explicit FDM can benefit from different levels of optimisations to fully exploit HPC systems.

\paragraph{Single-core level}
Algebraic operations and stencils calculations can benefit from basic compiler optimisations and code-vectorisation if the compiler has enough information and the data structures are adequately aligned and traversed.
Moreover, stencil calculations access neighbouring mesh point values from previous time-iterations, so data-blocking mechanisms enabling data reuse can improve data locality at the memory hierarchy.

\paragraph{Multi-core level}
Within explicit FDM time-steps, new values are computed from previous time iterations ones.
Hence, at each time step, the absence of data dependencies ensures the spatial loop's embarrassing parallelism. Such patterns can benefit from shared-memory parallelism through parallel programming model annotations. Explicit FDM codes must ensure balanced work partition and minimum synchronisation overheads.
Moreover, the neighbouring data dependencies across time-steps restrict the time loop parallelism, but the end of each time step is not globally synchronised; distant spatial regions can profit from asynchronous parallel time progress.

\paragraph{Cluster level}
The computation domain decomposition approach within explicit FDM corresponds to a spatial domain distribution across available nodes. Thus, each node is in charge of a portion of the initial domain, and data redundancy and message passing mechanism manage dependencies.
Such distribution must be scalable and preserve and combine with lower-level optimisations.

In this paper, we consider all previous points together and detail their application and combination.

\subsection{Saiph}
Saiph is a DSL easing the simulation of physical phenomena from the CFD domain in HPC environments.
Users specify CFD problems through high-level constructs defining systems of PDEs with time-space configurations.
Mathematical specifications, Saiph codes, evaluations and output results can be accessed online \cite{saiphURL}.

\begin{figure}
\centering
\includegraphics{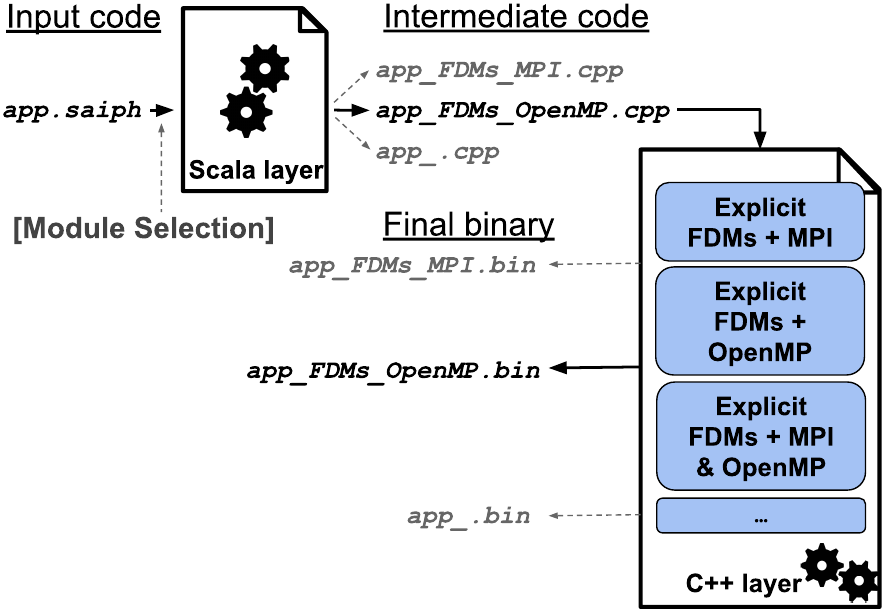}
\caption{Saiph layered abstraction}
\label{fig:SaiphDesign}
\end{figure}

The DSL is embedded in Scala \cite{scalavirt} and comprises two main divisions: the Scala and the C++ layer.
The Scala layer defines the language syntax; the C++ layer implements numerical methods and parallelisation strategies on separated libraries forming the modules.
Figure \ref{fig:SaiphDesign} illustrates the macroscopic compilation flow.
Input code is first compiled to be parsed by the Scala layer producing a C++ intermediate code.
The C++ code is then linked to the C++ library, and a second compilation produces the final parallel binary.
Because of this dissociation, discretisation methods and algebraic kernels are generic enough to support different input problem configurations.
Moreover, at the first compilation step, selecting a module determines the intermediate C++ generated code with function calls to the matching library.

At the Saiph intermediate output, PDEs are represented by equation trees modelled as abstract graphs. Vertices correspond to mathematical operators that relate the fluid fields designated by leaves.
At run-time, the numerical library traverses these graphs for each spatial coordinate at each time-step, applying basic parallelisation strategies.
However, single-core performance is far from optimal because graph data structures blind the compiler preventing optimisations.
This design offers high productivity and extensibility but limits Saiph from being a competitive HPC tool.
This paper enhances Saiph to leverage the specific domain knowledge and propagate it through the different compilation phases to achieve competitive and scalable performance.
We focus on generating a specific, detailed and optimisable spatial loop from the Scala layer for the input applications within the explicit FDM modules.
Moreover, we implement advanced parallelisation strategies at the underlying
C++ libraries, embracing and boosting the spatial loop execution, where computations and memory accesses happen iteratively.

\section{New Design and implementation}
We present a new design to extract and propagate information from the high-level code to the underlying layers.
Based on this information, we develop, adopt and combine different numerical parallel strategies.
As stated in the previous section, \ref{back}, the FDM-CFD algorithmic patterns occur within the spatial loop, at the mesh traversal.
To get a significant benefit in performance, we focus on optimising such a loop.

\subsection{Code generation}
\begin{figure}[h!]
\centering
\includegraphics{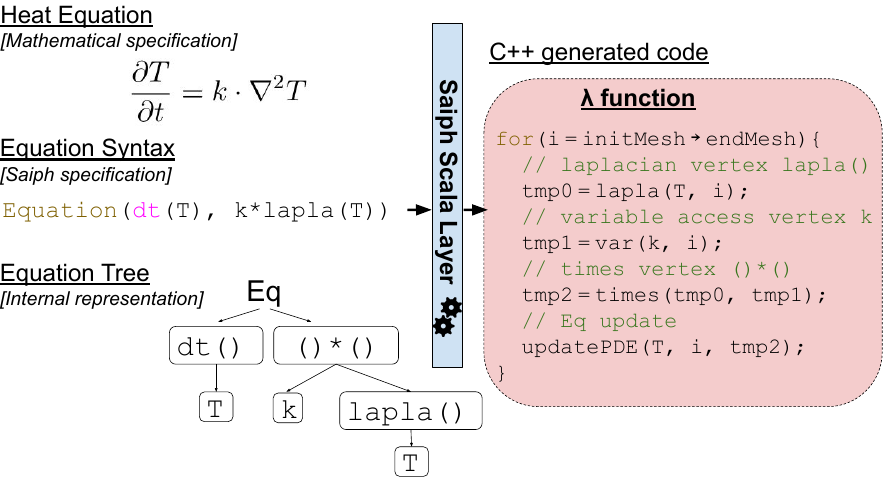}
\caption{Saiph equation specification, internal representation and code generation}
\label{fig:generatedEq}
\end{figure}

We present a generic build process ensuring productivity while enabling high-performance.
For an efficient equation resolution, we remove the spatial loop construct from the C++ library to generate it from the Scala layer encapsulating the generated code in a lambda function.
Figure~\ref{fig:generatedEq} shows how we generate a C++ lambda function from a high-level specification of a PDE.
From the input heat equation code, we traverse the equation tree at the Scala layer and generate, at the loop body, the corresponding method call for each vertex.
Computations and memory allocations and accesses remain at the C++ library, guaranteeing their unique, efficient implementation.

\begin{figure}[h!]
\centering
\includegraphics{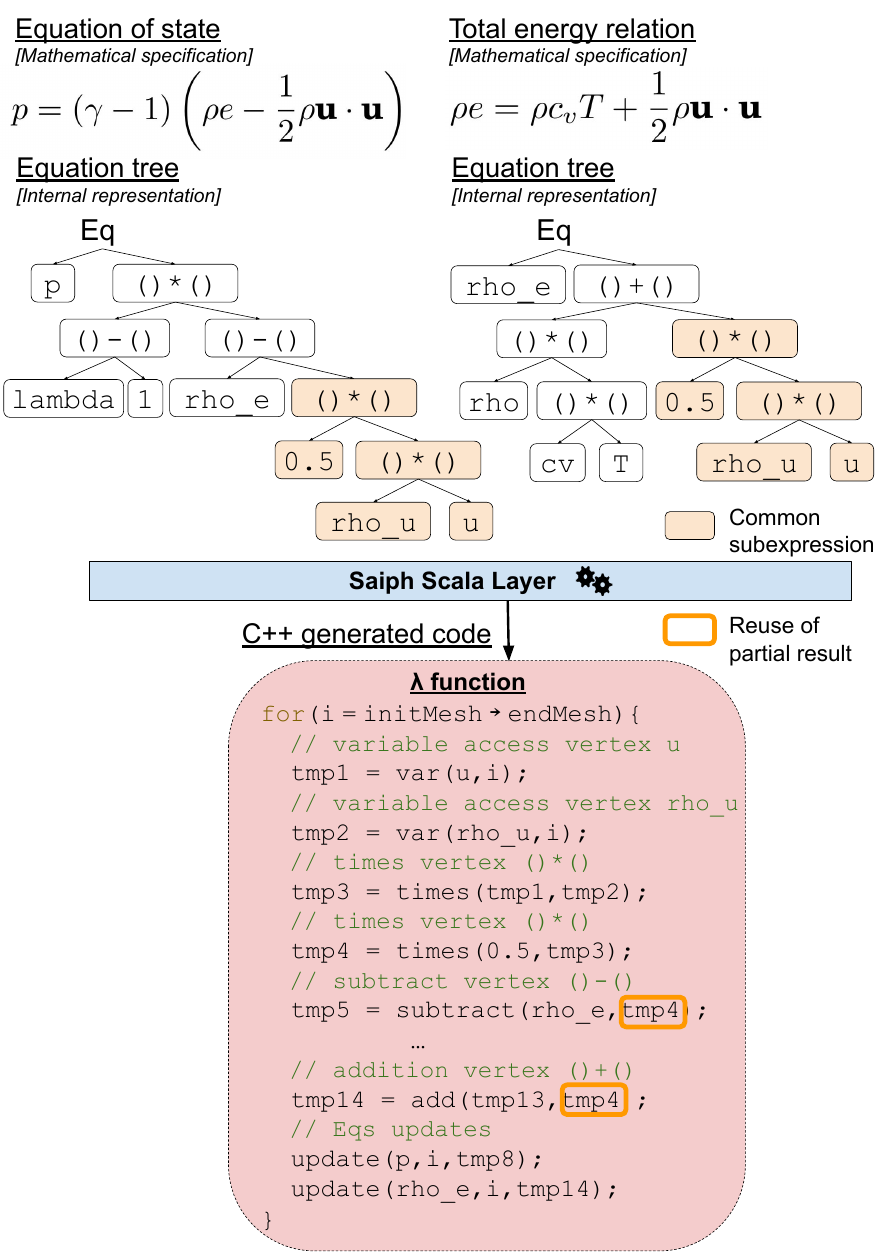}
\caption{Saiph partial results reuse at the equations code generation}
\label{fig:reuse}
\end{figure}

To minimise mesh traversals, we group the resolution of the equations into the same spatial loop.
Still, the equation final update depends on the equation's nature determined by the left-hand side expression.
We generate two independent lambda functions grouping the spatial resolution of time-derivative and non-time derivative equations, respectively.
For each loop body, we enable domain-specific optimisations, such as the common subexpression elimination (CSE) at the Scala layer through the Lightweight Modular Staging (LMS) \cite{lms:paper}, allowing partial results to be reused even for different equations.
Figure \ref{fig:reuse} exemplifies this reuse for two equations grouped into the same lambda; \texttt{tmp4} corresponds to the partial result of the common highlighted subexpression, so it is used at both equation updates.

\begin{figure}[h!]
\centering
\includegraphics{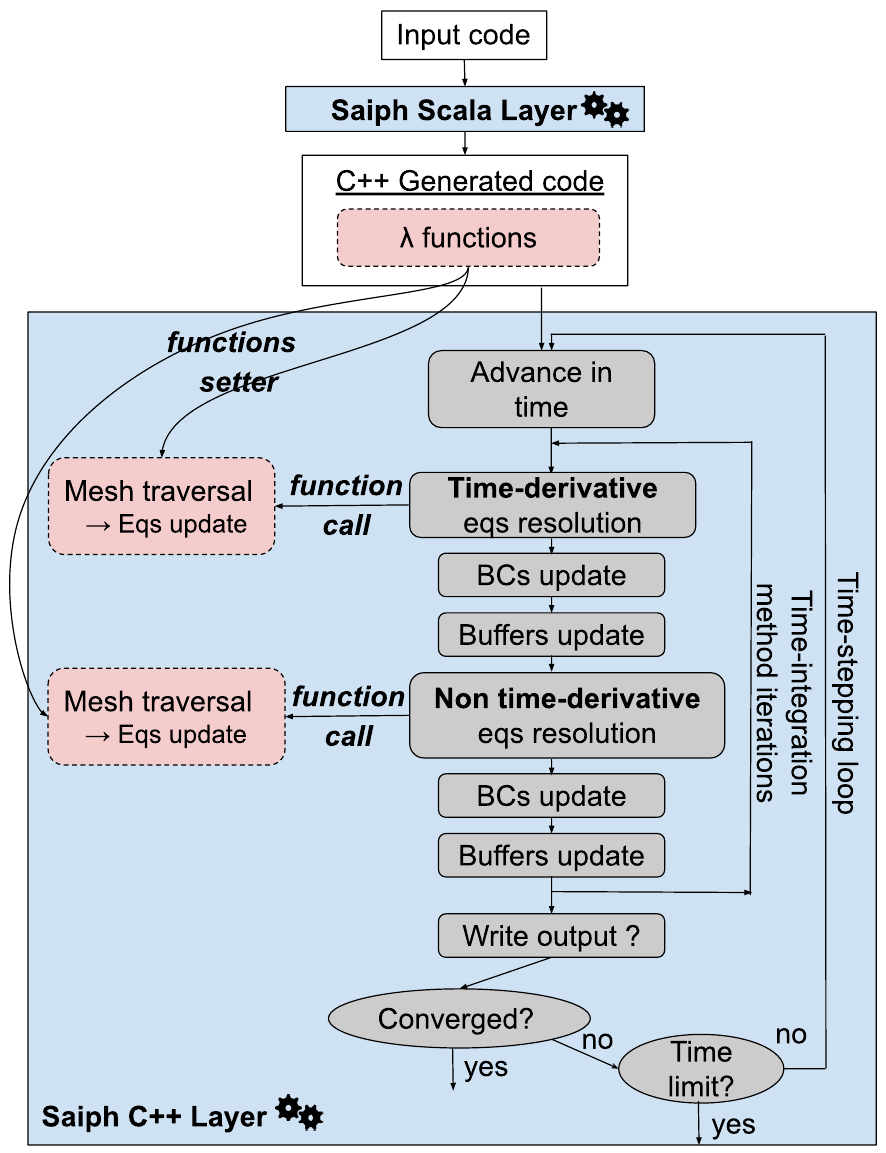}
\caption{Saiph resolution workflow}
\label{fig:workflow}
\end{figure}

Figure \ref{fig:workflow} illustrates the overall execution workflow.
Once the Scala layer outputs the C++ generated code from the input one, we use the \textit{lambdas} as arguments of new \textit{setters} methods from the C++ library, stating the \textit{resolution function} attributes.
Once these attributes are defined, we call them from the library within the integration methods' time loop.
The \textit{lambdas} executed at each time-step act as links between layers: from the generated code scope their capture the references of the fluid fields defined at the input code and call the C++ library methods that state the calculations over them.

Since we generate specific spatial loops automatically, we can further detail and annotate them to obtain the desired optimisation level.

\subsection{Exploiting low-level optimisations}~\label{section:lowLevelOpt}

\subsubsection{Code efficiency}
To benefit from compiler optimisations, we want detailed programs using simple structures and constant parameters.
The extraction of semantic information from the input code enables compile-time evaluations leading to an equivalent, more efficient program.
Linear algebra and stencil computations can enormously benefit from such transformations.
Within the spatial loop, Saiph equation resolution happens through function calls.
At the C++ layer, those functions perform variable accesses, algebraic operations, stencil computations and integration updates.
We implement them to encapsulate memory accesses and basic mathematical operations coded through simple structures involving conditional branches and few-iterations loops over problem variables.
Moreover, we define them as \texttt{static inline functions}, so these small, recurrent-used kernels avoid being called and escape the associated overhead.
We also generate the clause \texttt{\#pragma forceinline recursive} right before the spatial loop, at the \textit{lambda} function, to ensure the inline.
Consequently, we enable context-specific optimisations on the body of the inline functions.
By defining control variables such as mesh dimensions, operands dimensions and stencil neighbouring as literals, \texttt{const}, or \texttt{constexpr}, we expose them, and compile-time evaluation automatically applies.
To illustrate the internal implementation of such functions, Listing~\ref{VarOp} presents the function \textit{var}, for accessing a variable at a particular spatial position and returning its value. For that, the function receives four parameters: the pointer \texttt{varBuff} pointing the buffer at which the variable is stored, the integer \texttt{idx} specifying the spatial position to access, the integer \texttt{varDims} stating the dimensions of the variable and the pointer \texttt{res} indicating where to store the result.
Then, the \textit{var} function involves a loop that performs the desired variable access that can be unrolled if the variable's dimensions, \texttt{varDims}, are known at compile-time.

\begin{lstlisting}[frame=tb, numbers=none, basicstyle=\small, caption={C++ variable access kernel}, label=VarOp]{}
static inline real_t*  <@\textcolor{black}{var}@>(real_t* varBuff, int idx, int varDims, real_t* res) {
  for(int i = 0; i < varDims; ++i)
    res[i] = varBuff[idx*varDims + i];
  return res; }
\end{lstlisting}

\begin{figure}[h!]
\centering
\includegraphics{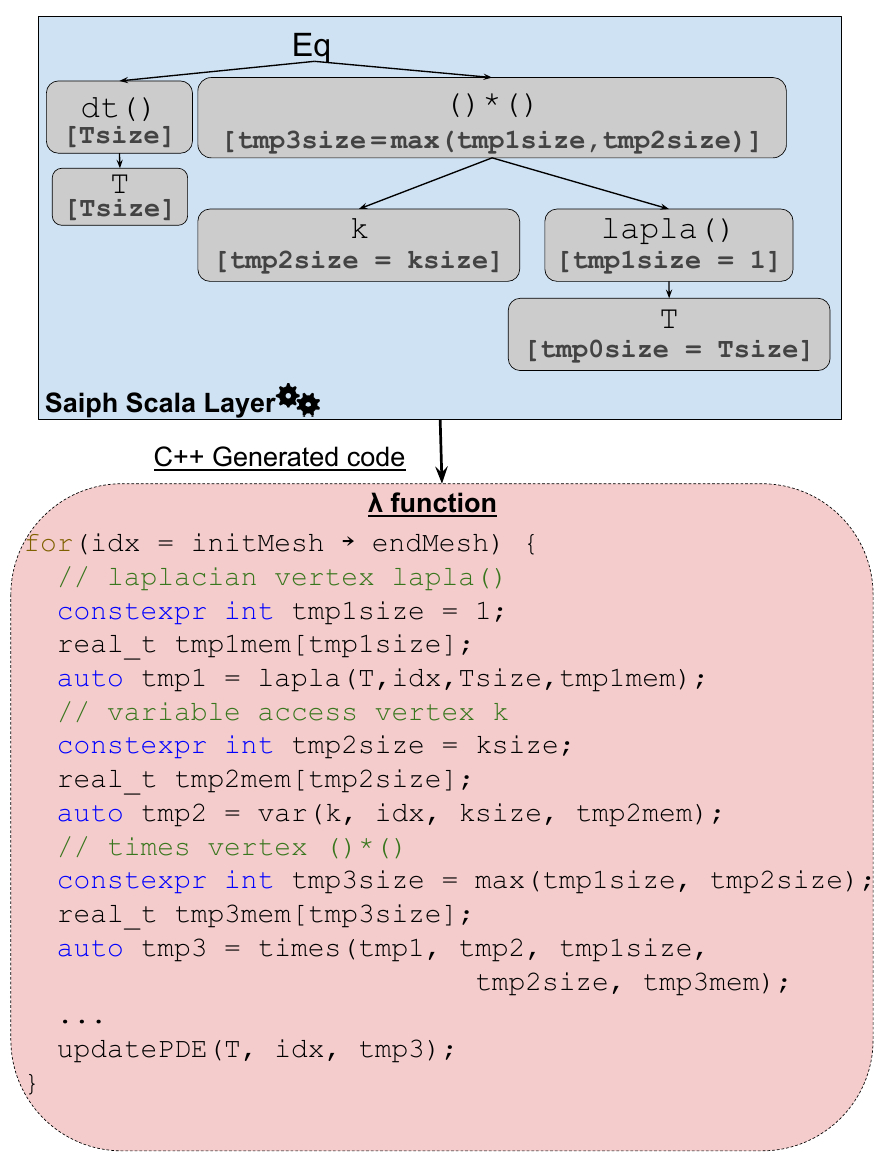}
\caption{Saiph equation characterisation and vertex operator generated calls}
\label{fig:HeatSize}
\end{figure}

Similarly, spatial derivative functions can be optimised since they iterate over mesh dimensions and stencil accuracy control variables, potentially derived at compile-time.
Hence, we want the Scala layer to generate an intermediate code with information over the control variables to benefit from compiler optimisations.
Mesh dimensions, fields dimensions and stencil accuracy are generated as literals by retrieving user code's information.
We extract the information for other kernels' control variables when traversing the equation tree.
At the bottom-up graph traversal, from the Scala layer, we calculate specific semantic information for each vertex from leaves' known dimensions, depending on the operator nature and children's dimensions.
We enhance the Scala layer to characterise graphs and generate kernel calls with evaluable control variables.
Figure \ref{fig:HeatSize} shows the tree characterisation of the heat equation and the specific generated \textit{lapla}, \textit{var} and \textit{times} function calls for computing a laplacian, accessing a scalar problem field and performing a product operation, respectively.
As we illustrate in the figure, we generate the operand dimensions' evaluation before the operator call and allocate the temporal memory to store the partial result.
Thus, the calls to the kernels are specific and determined at compile-time to be automatically optimised by the compiler.

\subsubsection{Micro-architecture use}\label{section:microArch}
\begin{figure}[h!]
\centering
\includegraphics{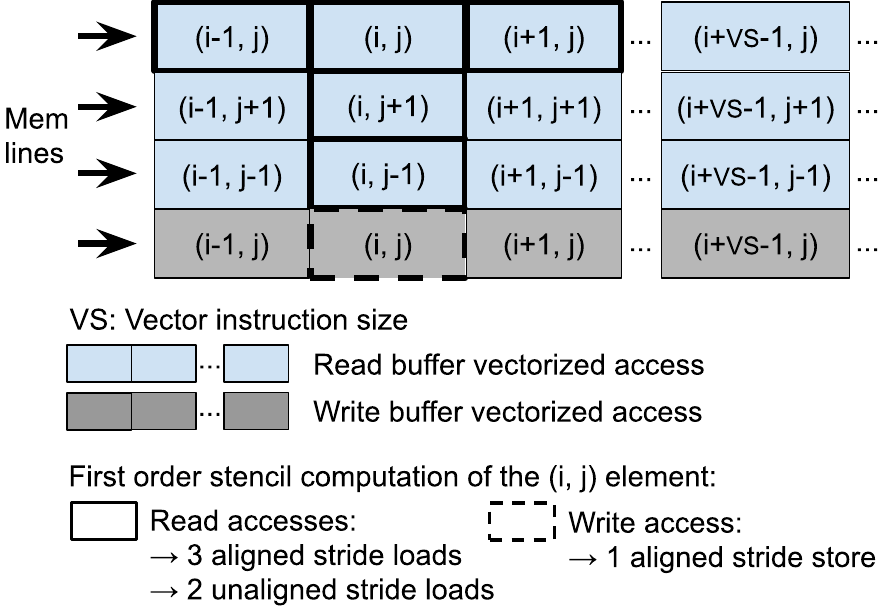}
\caption{Saiph memory access pattern for a first order stencil computation}
\label{fig:Stencil}
\end{figure}

As hardware moves forward to boost executions, we adapt code to the architecture design to well-exploit resources and core pipeline.
\paragraph{Vectorisation}
Due to its embarrassing parallel nature, the generated spatial loop can benefit from auto-vectorisation \cite{simd} through hardware support (SIMD length of vector units) and compiler loop dependence analysis.
Still, compilers need assistance in applying this optimisation: we leverage data alignment at the C++ library level for each field buffer by transparently allocating memory at an address multiple of parameter \texttt{alignment} set to the cache line size, optimal for memory movements.
Apart from base pointer alignment, vectorisation relies on known and aligned accesses.
Within Saiph FDM modules, stencil computations represent the principal cause of memory access.
Applying a stencil at a specific mesh element \texttt{(i, j, k)} implies accessing its neighbours.
For the first dimension, \texttt{nX}, the stencil involves contiguous memory accesses; the other dimensions neighbouring have non-unit strides, computed from the \texttt{(i, j, k)} element address plus an offset multiple of \texttt{nX}.
To maximise aligned accesses, we apply a padding strategy; \texttt{nX} adds extra points to force its size to be multiple of memory cache lines.
Thus, each data row perfectly fits into several cache lines ensuring contiguous or aligned stencil access.
Moreover, cache lines size is multiple of the vector units instruction size (set through \texttt{vectorSize} parameter) so, as long as the starting index of the spatial loop matches the start of a data row, the spatial loop is vectorised without peeled or remainder loops.
Figure~\ref{fig:Stencil} shows this organisation of data into memory and the automated aligned access pattern for a first-order stencil computation in a 2D spatial domain. The neighbouring access of the \texttt{(i, j)} element is contiguous for the first dimension and aligned for the second.
The padding strategy enables aligned and vectorised memory accesses.

To hint the compiler about data alignment, we automatically emit specific clauses at the intermediate code.
We declare field buffer pointers with \texttt{\_\_assume\_aligned} or \texttt{\_\_builtin\_assume\_aligned} attributes for Intel or GNU, respectively.
We mark \texttt{nX} and the lower bound of the spatial loop as multiples of \texttt{alignment}: \texttt{\_\_assume(nX\%alignment==0)} for Intel, \texttt{nX \& -alignment} for GNU, and we add the portable pragma \texttt{omp simd} right before the loop.
Similarly, we use the clause \texttt{omp declare simd} at library kernels to enable their SIMD versions.

\paragraph{Spatial blocking}
Stencil data-locality improves if implemented by means of data chunks; memory pieces fitting into the cache optimise their reuse \cite{tiling, tile}.
For that, we add blocking strategies at different levels of the DSL.
At the C++ library, once the mesh is discretised, we compute the number of local blocks $nbl_i$ per mesh dimension $i$ through iterative cuts based on the \texttt{L3size} parameter (L3 cache size).
We adopt a 2.5D blocking technique \cite{blocking1, blocking2}: blocking non-contiguous dimensions while streaming the computations over the first one ($nblX = 1, nblY \neq 1, nblZ \neq 1$).
The first dimension counts on the padding points, so the blocked spatial traversal maintains the innermost loop well-conditioned for vectorisation.
Once the number of blocks per dimension is set, we automatically derive the block sizes $bs_i$ parameters.
At the Scala layer, we generate a code with a generic spatial loop skeleton able to apply the blocking decisions that will take place at the C++ library.
Listing \ref{taskLoop} shows this generated code with the nested loops that enable the parametrised blocked spatial traversal; the three outer loops traverse the $nblZ*nblY*nblX$ blocks while the three inner loops traverse each block iteration space $bsZ*bsY*bsX$.
Preceding the loop, the generated code includes library method calls stating the loop bounds ($nblX$, $nblY$, $nblZ$, $bsX$, $bsY$, $bsZ$).

\subsection{Exploiting multi-core parallelism} \label{section:intraNode}

\subsubsection{Fork-join model}\label{section:forkJoin}
We transparently obtain a parallel code by generating the OpenMP clause \texttt{\#pragma omp parallel for collapse(3)} at the intermediate code, right before the embarrassingly parallel spatial loop.
The clause envelops the loop so that blocks are distributed among threads, each of them starting at a vectorisation beneficial aligned index.
Threads share the local memory, so we adapt the blocking decisions; we increment the number of blocks $nbl_i$ to have, at least, as many blocks as working threads (parameter \texttt{nThreads}) fitting simultaneously into the cache.
In such a way, there is enough parallel work to feed \texttt{nThreads} simultaneously while the blocks they compute in parallel fit into the L3 cache, occupying less than \texttt{L3size}, enabling good locality.

\subsubsection{Task model}\label{section:task}
\begin{figure}[h!]
\centering
\includegraphics{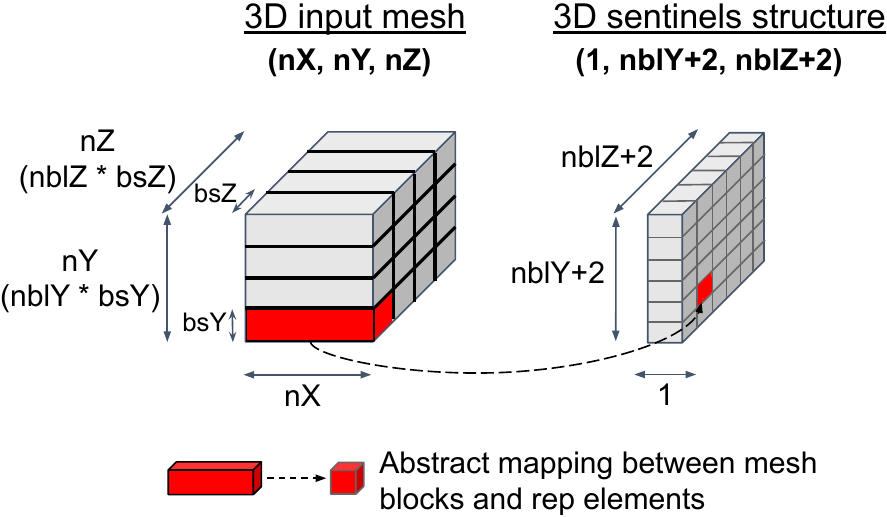}
\caption{Block mesh abstraction for the 2.5D blocking technique}
\label{fig:reps}
\end{figure}

This paradigm represents an appealing approach for HPC-CFD problems \cite{cfdTask, cfdTask2}.
The model requires annotating the code with task constructs enclosing pieces of code that will be asynchronously executed in parallel; we specify tasks data-dependencies to ensure correct execution order.
We develop two new DSL modules using OpenMP and OmpSs-2 \cite{pm:ompSs} programming models, respectively.
In both, we generate an annotated intermediate code creating tasks that envelop spatial block updates.
This code generation happening at the Scala layer precedes the allocation of problems fields.
Thus, to state task dependencies over the not-yet-allocated buffers, we use additional arrays whose elements are sentinels representing blocks.
At the C++ numerical library, we define such structure, shown in Figure~\ref{fig:reps}, as a multi-dimensional array of \texttt{chars}, with as many dimensions as the input mesh and as many elements as the number of blocks per dimension $nbl_i$  plus two.
Adding two elements per dimension allows us to specify stencil dependencies generically without worrying about boundary block cases.

We handle two of such structures mapping the source (read) and the destination (write) buffers, respectively, to state the time dependencies.
Finally, we use the pointers to the structures as lambda functions' arguments to enable their use at the spatial loop.
Listing~\ref{taskLoop} shows the annotated loop generated within the OmpSs2-module;
the skeleton of the spatial nested loops remains generic and we automatically generate a \textit{pragma} for the creation of tasks just before each block traversal.

\begin{lstlisting}[frame=tb, numbers=none, basicstyle=\small, caption={Generated spatial loop skeleton with OmpSs-2 annotations}, label=taskLoop]{}
for(int zb = 0; zb < nblZ; ++zb) {
  for(int yb = 0; yb < nblY; ++yb) {
    for(int xb = 0; xb < nblX; ++xb) {
      <@\textcolor{mauve}{\#pragma oss task in(repSRC[xb][yb][zb])}@> <@\textcolor{mauve}{\hspace{4pt}in(repSRC[xb+1][yb][zb])}@>
		            ...  <@\textcolor{mauve}{\hspace{5pt}out(repDST[xb][yb][zb])}@>
      for(int z = 0; z < bsZ; ++z) {
        for(int y = 0; y < bsY; ++y) {
         <@\textcolor{mauve}{\#pragma forceinline recursive}@>
         <@\textcolor{mauve}{\#pragma omp simd}@>
          for(int x = 0; x < bsX; ++x) {
\end{lstlisting}

\begin{figure}[h!]
\centering
\includegraphics{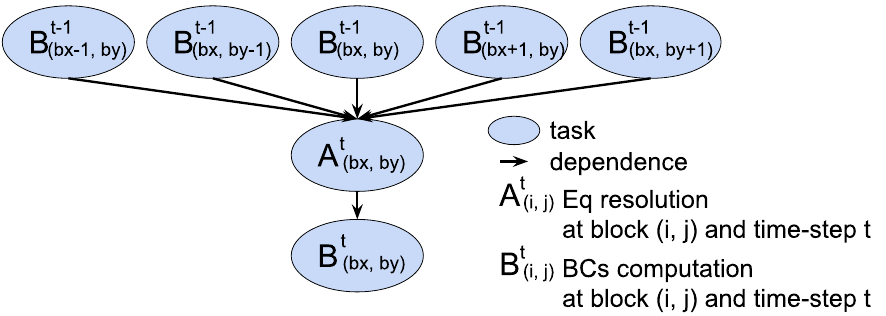}
\caption{Explicit FDM task scheme for a 2D block update}
\label{fig:tasksComm1}
\end{figure}

Tasks from the same time-step can be executed in parallel; hence, we adjust the blocking decisions to have at least \texttt{nThreads} blocks fitting simultaneously into the cache.
In such a way, the memory requirements from the tasks executed in parallel do not exceed the cache size, maintaining a good memory locality.
Block updates correspond to computational tasks, labelled as \textit{A-tasks} that satisfy the relation \textit{A-tasks}$= nblX*nblY*nblZ*niter$ where $niter$ is the number of time integration iterations of the simulation.
Moreover, taskifying the spatial loop requires protecting or applying the equivalent tasking strategy to other computations at the exact buffer locations.
Thus, we adapt BCs computations to happen within tasks, labelled as \textit{B-tasks}, at the same spatial blocks and use abstracted dependencies to relate them to the ones from the \textit{lambda} function.
We exploit the hidden temporal parallelism allowing asynchronous block time progress.
Figure \ref{fig:tasksComm1} shows the Saiph task scheme for a generic block update.
The task \textit{A} depends on previous \textit{B-tasks} over neighbouring blocks. Once executed, the corresponding \textit{B-task} is ready for execution.

\subsection{Exploiting distributed parallelism} \label{section:interNode}
We transparently apply a domain decomposition approach at the Saiph C++ library to exploit distributed parallelism~\cite{chung, cfdParallel}.
We partition the mesh to distribute a similar workload across the available MPI ranks.
For a highly scalable distribution, we enable cuts in every dimension so to create as many mesh portions as parameter \texttt{nMPI} which is the number of MPI ranks involved in the execution.
Naming $nbg_i$ as the number of blocks of the global mesh per $i$ dimension, and $nMPI$ the number of MPI ranks involved in the execution, our domain decomposition satisfies $nbgX*nbgY*nbgZ = nMpi$.
Hence, the number of cuts per dimension results from the factorisation of the $nMPI$ number.
We use the \textit{MPI\_Dims\_create} function~\cite{openMPI} to automate the computation of parameters $nbg_i$.
Alternatively, we can choose the proportions of the distributed blocks by manually setting the $nbg_i$ parameters.

\begin{figure}[h!]
\centering
\includegraphics{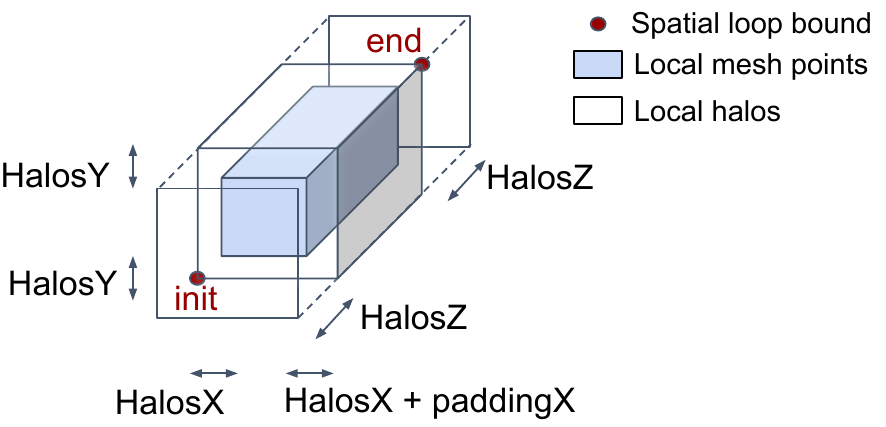}
  \caption{Distributed local 3D mesh with padding technique}
\label{fig:DistMesh}
\end{figure}

After each MPI rank receives its domain portion, we manage dependencies through data redundancy and message passing mechanisms.
Figure \ref{fig:DistMesh} schematises a local mesh portion of a 3D distributed mesh; local meshes are formed by mesh points partition and halos, which correspond to the replicated information from neighbours' local meshes.
We ensure a vectorisable contiguous spatial traversal while minimising redundant computations by traversing first dimension halos while avoiding last dimension halos updates.
Moreover, to adjust to the single/multi-core automated optimisations, we apply padding to the first contiguous dimension of each mesh portion.

We implement the communication of redundant data, or halos, to neighbouring processes at the end of each step.
Data is then automatically updated and correctly accessed at future computational steps.
We use the standard Message Passing Interface (MPI) library for communications.
Each MPI rank sends its updated frontier mesh points and receives halos boundaries from neighbour MPIs.
When distributing the mesh, we prioritise partitioning the last dimension to favour contiguous communication.
When cutting in more than one dimension, we apply gathering and scattering techniques for the non-contiguous halos messages.

\subsection{Exploiting hybrid parallelism} \label{section:HinterNode}
We mix multi-core and distributed parallelism within Saiph hybrid modules, combining fork-join/task model and message passing libraries to explore different levels of parallelism simultaneously.
We consciously combine them to preserve previous optimisations and obtain efficient configurations.

\begin{figure}[h!]
\centering
\includegraphics{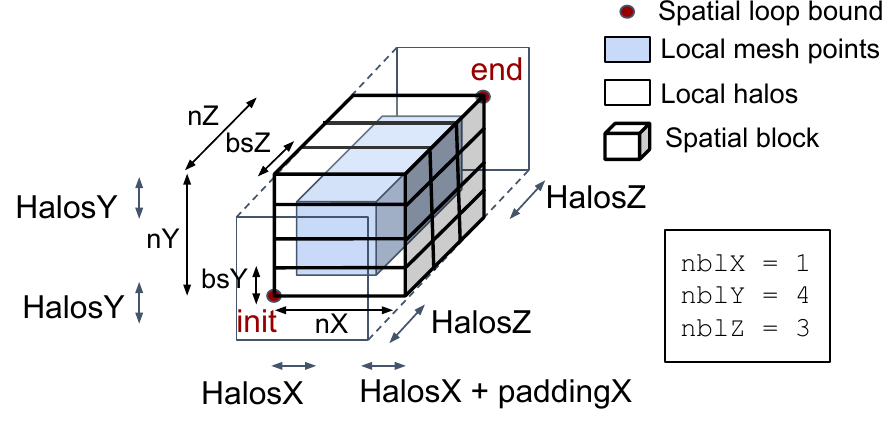}
\caption{Distributed local 3D mesh combining halos, 2.5D blocking and padding techniques}
\label{fig:blockDistr}
\end{figure}

For that, we apply the domain decomposition from Section~\ref{section:interNode} and distribute the mesh among the available MPI ranks. Locally, we group parallel computations into spatial blocks enabling the multi-core parallelism from Section~\ref{section:intraNode}. Moreover, blocking and padding strategies from Section~\ref{section:microArch} provide good memory usage and vectorisation.
Similarly to computations, we group boundary communication using the spatial block structure.
Figure~\ref{fig:blockDistr} illustrates such adjusted techniques over a distributed local mesh.

\begin{figure}[h!]
\centering
\includegraphics{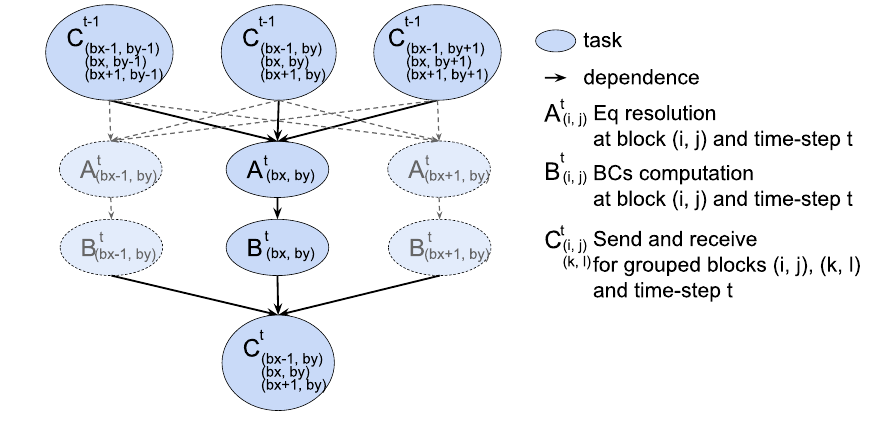}
\caption{Explicit FDM distributed task scheme for a 2D block update}
\label{fig:tasksComm2}
\end{figure}

Combining fork-join and domain distribution does not imply interoperability issues; we add a communication phase at the end of each time-step, where we exchange halos.
In contrast, combining tasks and domain partition require protecting communications using barriers or encapsulating them inside other tasks.
Similarly to mesh updates, we use the block structure from Figure~\ref{fig:blockDistr} to encapsulate the halos exchange into \textit{C-tasks}.
We internally state \textit{C-tasks} dependencies, for a correct execution order, through the same sentinels' structures from Figure~\ref{fig:reps}.
We use the Task-Aware MPI Library (TAMPI)~\cite{tampi} to favour the interoperability between MPI and OpenMP/OmpSs-2 by efficiently executing MPI operations from inside tasks constructs: the non-blocking TAMPI model avoids barriers and the underuse of computational resources while allowing the safe progress of the program execution.
Thus, computation \textit{A/B} and communication \textit{C} tasks are asynchronously executed in parallel whenever the dependencies are satisfied.
Figure~\ref{fig:tasksComm2} shows the task flow of a hybrid taskified time-step.

Mesh global and local blocking parameters ($nbgX$, $nbgY$, $nbgZ$) and ($nblX$, $nblY$, $nblZ$) are automatically derived by Saiph to balance computations among the available resources $nMPI$ and $nThreads$.
However, the default parameters can be modified, allowing the exploration of different computational granularities.
Regarding communication, we allow to group boundary block communication within the same \textit{C-task} to profit from established connections and thus, also enable communication granularity exploration, through parameters ($nCommX$, $nCommY$, $nCommZ$).

\section{Evaluation}
In this section we evaluate the quality of the code generated by Saiph and the parallelisation strategies implemented. On section~\ref{method} we describe the contextual framework for such evaluation. Then, sections~\ref{section:singleCore},~\ref{section:multiCore} and~\ref{section:Distr} conduct the Saiph performance evaluation at single-core, multi-core and cluster levels respectively.

\subsection{Methodology}\label{method}
\subsubsection{Saiph apps}
We use several Saiph applications~\cite{saiph:pasc} to evaluate our contributions. These applications are open-source~\cite{saiphURL}, and they include a reference output to validate their correctness.
Table~\ref{tab:appsDetails} details the applications used and their memory accesses and operations required to update a mesh point.
Each application involves a fixed number of stencil operations, \textbf{ops}, computed with a certain spatial accuracy, \textbf{acc}. This spatial accuracy is a Saiph parameter stated by users to determine the numerical precision of stencils. Computationally, the accuracy determines the number of neighbour points involved per stencil, which determines the computational intensity of the application.

\begin{table*}[!h]
\scriptsize
\centering
\caption{Saiph apps speedups and details per mesh point update depending on stencil operations and accuracy (\textbf{ops}/\textbf{acc})}
  \label{tab:appsDetails}
  \addtolength{\leftskip} {-1cm} 
  \addtolength{\rightskip}{-1cm}
  \begin{tabular}{lcccc}
    \toprule
    App & Mesh points \&  & Mem access/point & FLOP/point & Saiph\textit{-lambda} \\
		& Mem. use (MB) & $x+$\textbf{ops}(\textbf{acc}+1) & $y+$\textbf{ops}(2( \textbf{acc}+ 1)+1) & vs \textit{tree-traversal}\\
\midrule

    1DSineWave  & 100001 - 1.5 & 13+1(\textbf{acc}+1) & 2+1(2(\textbf{acc}+1)+1) & 3,9x\\
    2DSmithHutton  & 2001*1001 - 61 & 24+4(\textbf{acc}+1) & 8+4(2(\textbf{acc}+1)+1) & 13,9x\\
    2DInviscidVortex  & 1601*1601 - 215 & 69+12(\textbf{acc}+1) & 25+12(2(\textbf{acc}+1)+1) & 13,8x\\
    3DHeat  & 301*201*201 - 185 & 13+3(\textbf{acc}+1) & 2+3(2(\textbf{acc}+1)+1) & 4,5x\\

\bottomrule
\end{tabular}
\end{table*}

\subsubsection{Hand-tuned codes}\label{section:handT}
We develop hand-tuned codes for each of the Saiph applications tested. Such codes explicitly implement the targeted single-core optimisations that Saiph automatically applies.
We validate the results produced by the optimised hand-tuned codes and we use them as baselines to appraise the performance of Saiph automatic optimisations.

\subsubsection{Yask kernels}
Yask \cite{yask} is a state-of-the-art framework for the exploration of the HPC stencil-performance design space.
The tool provides automatic stencil optimisations, including cache blocking, vector folding and vectorisation from high-level stencil kernels specifications and generates, from the user code, parallel programs using multiple cores through OpenMP threads.
Yask and Saiph tackle different domains and levels of abstractions but can be compared regarding the quality of the code generated. To conduct such a comparison, we develop a 3D Heat equation kernel within Yask. For that, we use Yask version 3.05.06 \cite{yaskV} and compile the kernel using different compilers with default optimisation parameters.

\subsubsection{Tests}
All applications use double-precision floating-point formats.
We firstly address, on Section~\ref{section:singleCore}, the single-core absolute performance of Saiph generated code using different native compilers and compare such results against manually optimised codes and literature results. Then, on Sections~\ref{section:multiCore} and~\ref{section:Distr}, we explore parallelisation strategies through scaling performance studies tackling the three levels of HPC resources.

This evaluation procedure aims to validate the proposed collection of optimisations and the strategies to apply and combine them automatically to demonstrate the DSL scalability and high performance.
Quantifying productivity and comparing Saiph to other CFD tools is an extremely challenging task because each framework targets a different level of abstraction and provides different optimisations.
This comprehensive evaluation escapes the scope of the paper, and we contemplate it as future work.

\subsubsection{Hardware}
We run the applications on BSC's Marenostrum 4 supercomputer \cite{mn4}.
Compute nodes are equipped with two sockets, Intel Xeon Platinum 8160 CPU with 24 cores each, supporting vectorisation instructions up to AVX-512 and sharing an L3 cache of 33MB.
The experiments in Section~\ref{section:multiCore} are limit to single socket to avoid NUMA issues. For the same reason, the hybrid MPI+OmpSs-2 experiments described on Section~\ref{section:Distr} use one MPI rank per socket. In this case, the 24 cores assigned to each rank are exploited through OpenMP/OmpSs-2 threads.

\subsubsection{Parametric runs}
Saiph runs are defined by the set of parameters (\texttt{alignment}, \texttt{vectorSize}, \texttt{L3size}, \texttt{nThreads}, \texttt{nMPI}, \texttt{commBlocks}).
The firsts are hardware-dependent, and we set them to 64 bytes (L3 lines size), 8 (double-precision floating-points fitting on AVX-512) and 33MB, respectively.
Regarding computing resources, we select \texttt{nThreads} and \texttt{nMPI} according to the execution test.
Finally, the \texttt{commBlocks} parameter allows exploring performance enhancement at hybrid TAMPI and OmpSs-2 runs. We empirically set it to 4.
Once set, parameters automatically combine and determine the code transformations that lead to specific optimised parallel code.

\subsubsection{Compilers}
Table~\ref{tab:comp} displays the compiler versions and corresponding flags used to compile Saiph, Yask and the hand-tuned codes. All the codes involving the OmpSs-2 programming model have been compiled using LLVM; for other codes, if nothing is specified, binaries are built from Intel.

\begin{table}[!h]
\centering
\caption{Saiph intermediate compilers and flags}
  \label{tab:comp}
  \begin{tabular}{ll}
    \toprule
    Compiler-Versions & Flags \\
\midrule
    ICC - 2020.1 & \texttt{-O3 -qopenmp -qopenmp-simd} \\
    \textit{icpc}& \texttt{-inline-forceinline} \\
                 & \texttt{-xCORE-AVX512} \\
                 & \texttt{-qopt-zmm-usage=high}\\
\midrule
    GCC - 9.2.0 & \texttt{-O3 -fopenmp -fopenmp-simd} \\
    \textit{g++}& \texttt{--forceinline -ftree-vectorize} \\
                & \texttt{-march=skylake-avx512}\\
\midrule
    LLVM- 13.0.0          &  GCC flags \\
    ompss-2\cite{pm:llvm} & [+] \texttt{-fompss-2}\\
    \textit{clang++}      & \\
\bottomrule
\end{tabular}
\end{table}

\subsection{Single-Core Performance}\label{section:singleCore}
We test the performance of spatial loops for the applications of Table~\ref{tab:appsDetails} using a default stencil accuracy of 2. At this level, we choose the small problem sizes from Table~\ref{tab:appsDetails} to stress vector units and assert the generated code's effectiveness.

\subsubsection{Saiph-\textit{lambda} vs Saiph-\textit{tree-traversal}}
We start comparing the new method to solve Saiph's equations based on \textit{lambdas'} generated at compile-time against the original method based on traversing equation trees at runtime.
Results are presented in the last column of Table~\ref{tab:appsDetails}.
Saiph based on \textit{lambdas} outperforms old implementations \cite{saiph:rwdsl, saiph:pasc} by a factor of 4x to 14x depending on the length of the equation trees since the new implementation avoid traversing it at each time step.
This performance increase comes from the fact that lambda functions generated at compile-time enable native compiler optimisations that cannot be applied when tree-traversals take place at runtime.

\subsubsection{Hand-tuned code vs Yask kernel}
To evaluate the quality of our hand-tuned codes introduced in section~\ref{section:handT}, we compare the single-core performance of the 3D Heat equation application manually implemented against Yask results.
The application mainly involves computing a second-order stencil, which can be easily specified using Yask.
Moreover, stencils represent the most challenging patterns to optimise within Saiph. Thus, by proving the high quality of stencil computations, we demonstrate the overall quality of the explicit FDMs for CFD.
We confront the 3D Heat equation hand-tuned code against the corresponding Yask kernel, using different compilers. Table~\ref{tab:handVsYaskVsSaiph} show the performance results of the different implementations of the 3D Heat equation running on a single-core. Results are reported using memory bandwidth, GFLOPS and Mpoints/s as absolute performance metrics. While the first two are implementation and optimisation dependent, Mpoints/s is based on the problem size, a user-fixed parameter. Hence, we focus on this last metric for a more fair comparison across implementations.

\begin{table}[h!]
\centering
\caption{Single-core performance comparison between different implementations of the Heat3D application}
  \label{tab:handVsYaskVsSaiph}
  \begin{tabular}{lccccc}
    \toprule
Compiler    \ & Heat3D app     & GB/s & GFLOPS & \textbf{Mpoints/s}\\
\ & implementation &      &        &         \\
\toprule
    \multirow{3}{1mm}{icpc} & Hand-tuned & 99.01 & 12.87 & \textbf{574} \\
                           & Yask       & 61.41 & 9.92  & \textbf{524} \\
                           & Saiph      & 96.43 & 12.54 & \textbf{559} \\

\midrule
    \multirow{3}{1pt}{g++} & Hand-tuned & 32.79 & 4.26 & \textbf{190} \\
                           & Yask       & 24.25 & 3.90 & \textbf{230} \\
                           & Saiph      & 27.82 & 3.61 & \textbf{161} \\

\midrule
    \multirow{3}{1pt}{clang++} & Hand-tuned & 97.09 & 12.62 & \textbf{562} \\
                            & Yask       & 58.92 & 9.49  & \textbf{586} \\
                            & Saiph      & 98.14 & 12.76 & \textbf{569} \\

\bottomrule
\end{tabular}
\end{table}

Hand-tuned and Yask implementations deliver comparable results between each other and when using Intel and LLVM compilers. However, when using the Gnu C++ compiler, the Yask kernel runs about 56\% slower than the same kernel built with the Intel C++ compiler. This performance drop is higher for the hand-tuned implementation, which shows a 67\% of performance drop when using \textit{g++}.
Overall, those results demonstrate that our manually developed codes successfully encode optimisations benefiting explicit FDMs patterns. Moreover, the compiler choice determines the performance of the final binary.

\begin{figure*}[!h]
\centering
\includegraphics[width=1\textwidth]{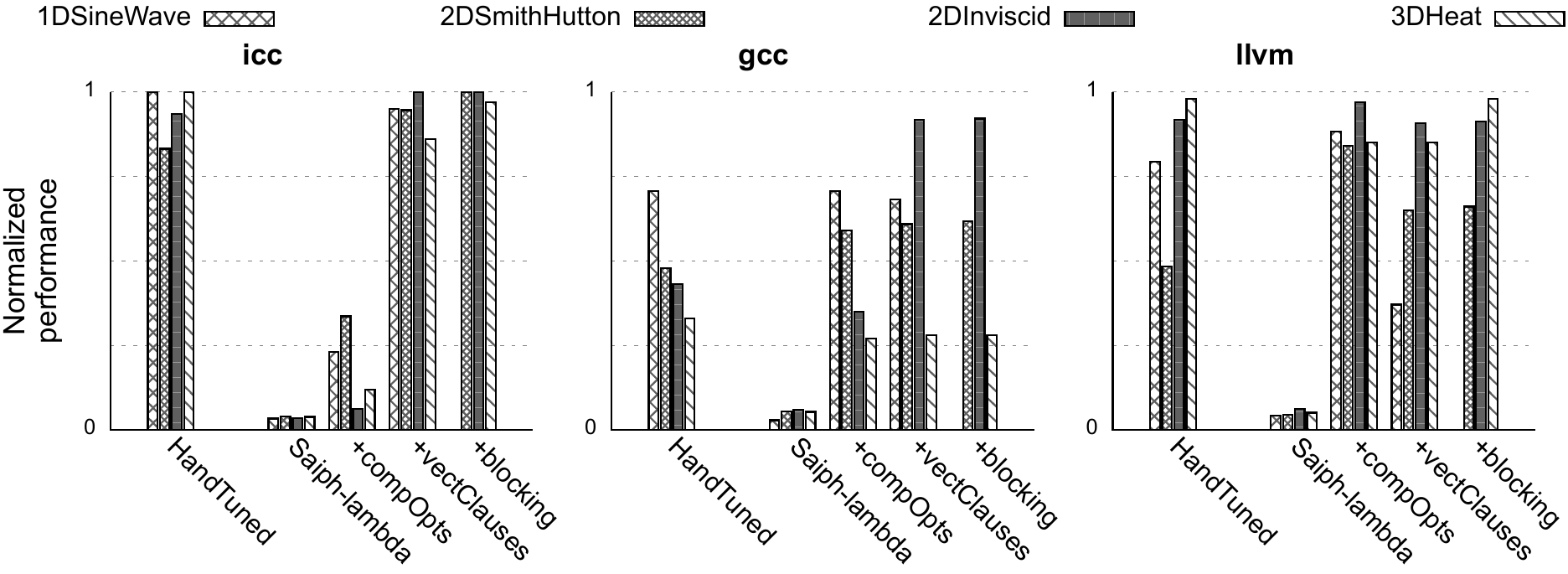}
\caption{Normalised performance of single-core CFD spatial loops executions}
\label{fig:singleCore}
\end{figure*}

\subsubsection{Saiph vs hand-tuned codes}
We appraise Saiph automated optimisations from implementations in Section~\ref{section:lowLevelOpt} against our manually optimised codes versions, using different native compilers.
Figure~\ref{fig:singleCore} shows normalised results, taking the most performant run as the baseline for Intel, GNU and LLVM compilers.
At each barplot, the first set of bars corresponds to hand-tuned loop bodies.
The second set, \textit{Saiph-lambda}, refers to the new implementation based on the lambdas, described above, which do not include any additional optimisation. We use this version as the baseline for the last three code sets, incrementally enabling compile-time evaluations, vectorisation and blocking.
For \texttt{icpc}, the most optimised Saiph version provides 28x, 25x, 28x, and 26x of performance increase compared to the \textit{Saiph-lambda} results and 0.95x, 1.2x, 1.1x, and 0.98x against hand-tuned codes.
The use of other compilers shows similar results.
Although the \texttt{g++} compiler presents lower performance results overall, Saiph optimised codes reach or surpass hand-tuned ones' performance.
Such results illustrate how compilers can produce a more efficient binary when optimising the regular loop body structures from the Saiph generated code than when receiving our hand-tuned constructs.
Saiph generated codes appear to be more optimisation-friendly for some applications and compilers than hand-tuned codes developed as human programmers.
Moreover, binaries built from \texttt{g++} and \texttt{clang++} show how vectorisation can be enabled without clauses, happening at \textit{+compOpts} versions.
In such cases, we see a slight performance drop when using clauses: compiler transformations can produce different codes depending on the order they are applied so that default automated decisions can be preferable.
However, benefits, when such transformation is not guaranteed, justify the use of clauses.
Finally, multi-dimensional blocking shows low impact because substantial concurrency is necessary to push the limits of the memory system \cite{memlimit}.
However, memory pressure at the cache level will appear for memory demanding cases under shared-memory parallelism.

\subsubsection{Saiph vs Yask}
Finally, we compare Saiph single-core optimised code against Yask results. For that, we use the 3D Heat equation application and the fully automatically optimised Saiph version. Table~\ref{tab:handVsYaskVsSaiph} adds Saiph performance results along with the already evaluated hand-tuned and Yask ones.
Numbers illustrate how Saiph single-core performance is comparable with Yask results. Using the GNU C++ compiler, Saiph suffers from a higher drop in performance than Yask, compared to Intel results. Binaries built from other compilers show competitive performance.
Saiph single-core optimisations and their automatic application are thus validated.

\subsection{Multi-core Scaling Performance} \label{section:multiCore}
\begin{figure}[h!]
\centering
\includegraphics[width=1\textwidth]{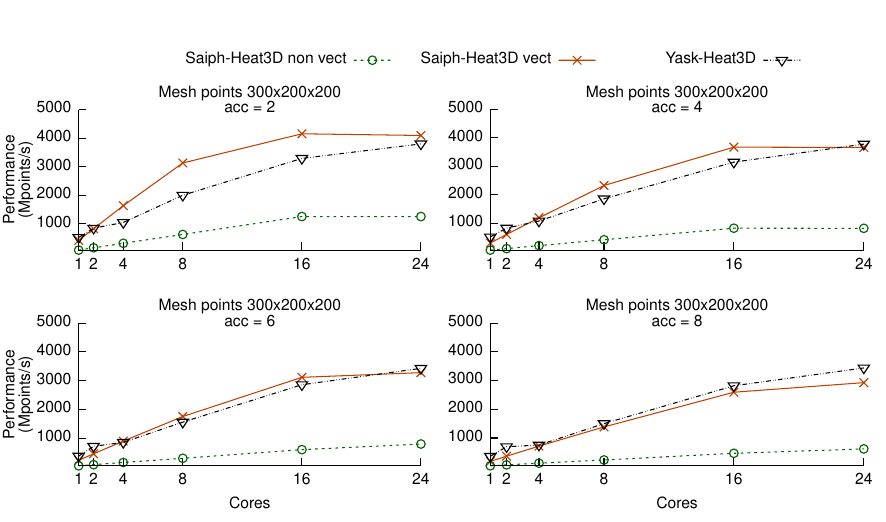}
\caption{Multi-core performance comparisons of Saiph and Yask OpenMP fork-join 3D Heat application for several spatial accuracy orders}
\label{fig:multiCoreRoof}
\end{figure}

We use the already evaluated Saiph optimised single-core results as baselines to address the spatial loop's scalability at the multi-core level.
We evaluate the fork-join model from section~\ref{section:intraNode} and verify that previous optimisations are preserved.
Figure~\ref{fig:multiCoreRoof} shows results up to a socket (24 cores) for the 3D Heat application from Table~\ref{tab:appsDetails} comparing Saiph vectorised and non-vectorised implementations against Yask results for different spatial accuracy orders.

Figure~\ref{fig:multiCoreRoof} shows decreased performance at rising space order for all the codes tested. Moving to higher accuracy implies increasing the stencil length, leading to more computations for every point update.
Using the Heat application and moving from a second to an eight space order corresponds to a theoretic computer demand increase of $2.6\times$ from 23 FLOP/point to 59 FLOP/point.
Saiph vectorised results for the 1-core executions give 420 Mpoints/s and 183 Mpoints/s at $acc=2$ and $acc=8$, respectively, representing a $2.3\times$ of performance decrease.
Hence, the Saiph automated combination of low-level optimisations apply for different space orders.
Moreover, Saiph vectorised results from Figure~\ref{fig:multiCoreRoof} outperform scalar ones by a factor of 4, taking advantage of the vector instruction size.
Finally, Saiph's results are competitive compared to Yask's ones.
Nevertheless, Yask optimisations give slightly better performance at high spatial accuracy orders.

We conduct a roofline analysis~\cite{roofline} using the Intel Advisor Roofline tool~\cite{intelRoof} to obtain computational performance, and memory bandwidth values for the Saiph vectorised runs using 16 cores and a second-order spatial accuracy.
Within such runs, the spatial loop delivers 742.88 GB/s, surpassing the DRAM and L3 bandwidth peak performances of 114 GB/s and 353 GB/s, respectively.
Stencil computations are usually memory-bounded, but the working set of the analysed application is small ($\sim185 MB$). Moreover, the Saiph blocking technique enables the reuse of cached memory, crucial for surpassing the roof of L3 cache bandwidth.
Hence, the loop's performance is not limited by the simultaneous accesses of the 16 cores to the same L3 cache of 33 MB.
A deeper analysis characterises the loop as a cache-bound workload, where the L1 and L2 cache stalls are the most significant cause of performance loss.
Saiph optimised results are close to the machine peak performance.
The Saiph automated combination of low-level optimisation and shared-memory parallelism is then validated at different space orders. Saiph multi-core support provides competitive performance close to Yask' and the processor peak performance.

\begin{figure}[h!]
\centering
\includegraphics[width=1\textwidth]{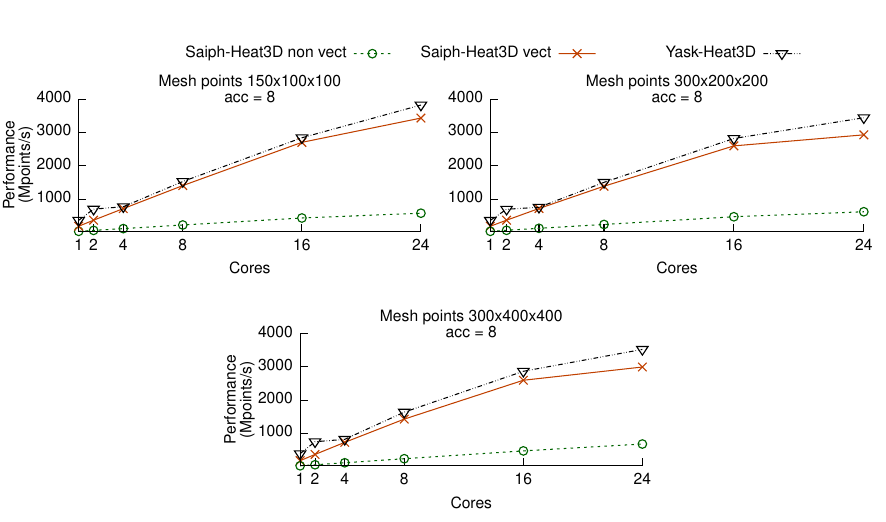}
\caption{Multi-core performance comparisons of Saiph and Yask OpenMP fork-join 3D Heat application for several mesh sizes}
\label{fig:multiCoreRoofMesh}
\end{figure}

In order to evaluate the impact on performance of the problem size, we perform several executions running the Heat application with different mesh sizes: 150x100x100 points occupying $\sim24$ MB, 300x200x200 points occupying $\sim185$ MB and 300x400x400 points occupying $\sim738$ MB.
Results are presented in Figure~\ref{fig:multiCoreRoofMesh}.

Figure~\ref{fig:multiCoreRoofMesh} shows slightly better results for the minor mesh size fitting into the L3 cache.
Saiph outputs similar results for bigger meshes, validating the efficacy of Saiph's multi-dimensional blocking strategy under shared-memory parallelism. Again, Saiph vectorised results are competitive with Yask's ones.

\begin{figure}[h!]
\centering
\includegraphics{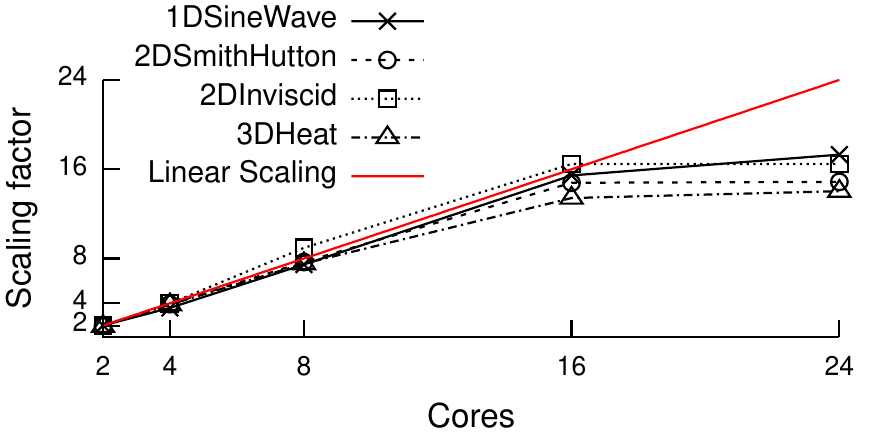}
  \caption{Fork-join parallel scalability at socket level}
\label{fig:multiCoreScaling}
\end{figure}

We repeat the previous scaling study for the different Saiph applications from Table~\ref{tab:appsDetails} using a second-order spatial accuracy.
Figure~\ref{fig:multiCoreScaling} shows scaling results for blocked and vectorised applications taking the optimised single-core execution as a baseline.
We add a linear scaling curve with an arbitrary origin value to have the slope reference for linear scalability.

\begin{figure*}[!h]
\includegraphics[width=1\textwidth]{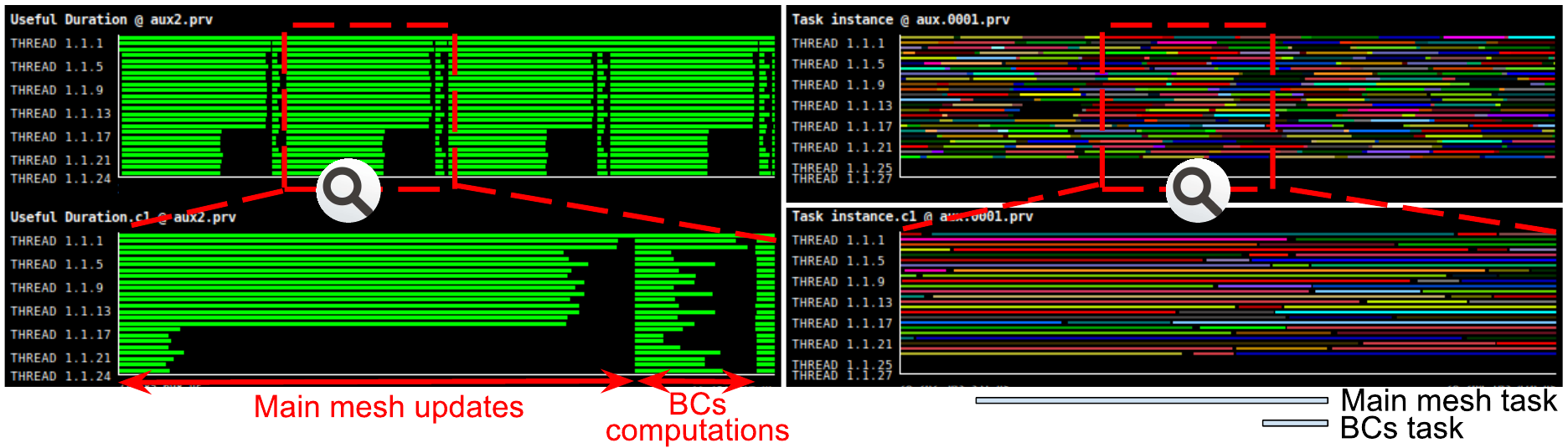}
\caption{Main mesh and BCs updates using 24 threads and fork-join (left) and tasks (right) models}
\label{fig:multiCoreTraza}
\end{figure*}

All runs from figures~\ref{fig:multiCoreRoof} and~\ref{fig:multiCoreRoofMesh}, scale up to 16 cores due to load imbalances at the 24-cores executions.
We assess that Saiph blocking technique ensures good memory locality and enough parallel work for linear scaling performance. However, load imbalances at the 24-cores executions stall the scalability as the number of threads is not multiple of the number of local blocks.
The number of cores \texttt{nThreads} determines the blocks, so the performance of the parallel code is suboptimal for ill-suited configurations;
load imbalances appear if \texttt{nThreads} is not multiple of the number of blocks.
This fact can be hypothesised when using 24 threads and confirmed with the corresponding trace on the left of Figure~\ref{fig:multiCoreTraza}.
The execution on the right of Figure~\ref{fig:multiCoreTraza} shows how using the tasking module can prevent such a performance drop.

The shared-memory parallelism is firmly bound to the automatic loop blocking technique and the balanced work partition, leading to linear scalability under the appropriate combination of parallel paradigms and hardware resources.
Saiph eases the exploration of such key parameters that affect performance, giving insight into the best running configurations. In the previous tests, we realised that runs using 16 cores provide a similar performance as those using 24 cores but using less power. Similarly, Saiph modules permit us to compare different parallel paradigms to choose the best fitting one.

\subsection{Distributed Scaling Performance} \label{section:Distr}
Lastly, we evaluate the applications' distributed parallelism through weak and strong scaling studies.
For that, we use bigger problem sizes and evaluate memory bound workloads and communications overheads.

\begin{figure}[h!]
\centering
\includegraphics{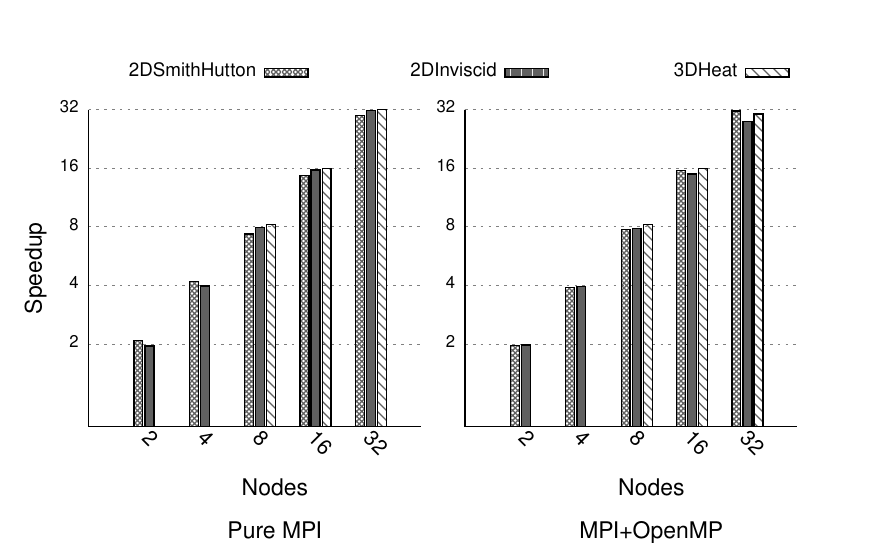}
\caption{Weak scaling using different parallel strategies}
\label{fig:Weakscaling}
\end{figure}

We present a weak scaling study for the evaluation of the distribution strategy from sections~\ref{section:intraNode} and~\ref{section:HinterNode}.
For distributed executions assessment, communications and required data manipulations play an essential role.
Thus, we take one-node executions (2 MPIs) as baselines for 2D applications and four-node executions (8 MPIs, 3D mesh partition) for 3D use cases.
As illustrated in Figure~\ref{fig:Weakscaling}, the multidimensional applications of Table~\ref{tab:appsDetails} present a linear scalability of up to 32 nodes for pure MPI and hybrid MPI+OpenMP fork-join modules.

\begin{figure}[h!]
\centering
\includegraphics{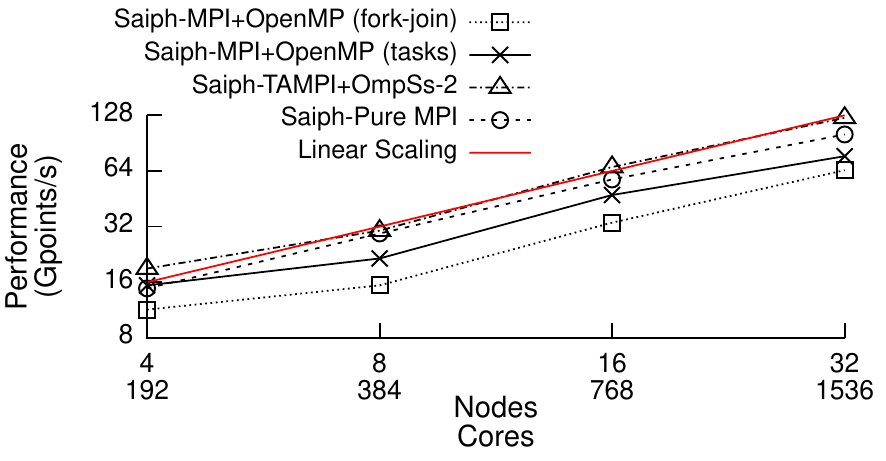}
  \caption{Strong scaling for the 3D Heat application using different hybrid parallelisations}
\label{fig:Strongscaling}
\end{figure}

Finally, we carry out a strong scaling study for the 3D Heat equation application.
For that, we enlarge the spatial mesh, set the spatial stencil accuracy to 8 and compute 500 time-steps. According to Table~\ref{tab:appsDetails}, the new problem size of 2401*1201*1201 mesh points, occupy 51.6 GB (2 buffers read/write using double precision) and involves 63 Tera memory accesses, 4.72 Tera stencils and a total of 93 TFLOPs.
Figure~\ref{fig:Strongscaling} displays the absolute performance behaviour from 4 to 32 nodes using and combining different shared-memory and distributed parallel paradigms from Saiph.
We add a linear scaling curve with an arbitrary origin value to have the slope reference for linear scalability.
Pure MPI runs use as many MPI processors as available cores in the nodes, while hybrid runs bind MPI processors to sockets and use as many threads as cores in the socket.
Finally, the blocking technique is automatically applied to each local mesh to use the most appropriate granularity for the multi-core parallel work.

Distributing the same problem size across a different number of nodes illustrates the transition from memory-bound to compute-intensive workloads ending up in scenarios where communications overheads become costly.
Saiph hybrid executions show linear scaling up to 16 nodes.
When using 32 nodes, local computations decreases while communications increase, producing imbalances leading to a non-linear performance scaling.
Overall, task versions deliver higher performance and better scalability than the fork-join version by reducing computation load imbalances.
Concretely, the use of TAMPI presents higher performance than the rest of Saiph versions, including the pure MPI one.
TAMPI allows computations and communications to overlap, increasing the overall performance and scalability of the runs.
By taking the four-node execution as a reference, the TAMPI-OmpSs-2 module outperforms the MPI fork-join, tasks and pure versions by a factor of 1.7x, 1.3x and 1.3x, respectively.
Those results validate Saiph implementations and demonstrate its competitive results. Different modules allow a parallel exploration to select the most performant paradigms.

\section{Related Work}
Many tools target the CFD domain to deliver performance, portability, or productivity \cite{CFDapps}.
Mainly divided into two main branches, there are frameworks tackling stencil computation performance and DSLs facing complete CFD, PDEs system resolution.

In the first group, different existing frameworks provide automated low-level stencil optimisations and parallel implementations.
Liszt \cite{liszt} is a DSL for unstructured mesh computations. The language is designed for code portability across heterogeneous platforms.
Liszt users work at the numerical level giving information to ensure that the compiler can infer data dependencies.
Similarly, for unstructured mesh computations, OPS/OP2 \cite{ops, op2} and PyOP2 \cite{pyop2} as the Python extension, give an abstraction for stencil computations at CPUs, GPUs, and distributed systems. Such DSLs are embedded in C/Fortran and Python, respectively.
Yask \cite{yask} is a C++ library also for automatic stencil optimisations. Cache blocking, vector folding and vectorisation are automatically applied. Moreover, Yask generates from the user code parallel programs using multiple cores and distributed-memory parallelism.
ExaSlang \cite{exaslang} is another stencil-specific programming language that provides different layers of abstraction and exploits domain information for neighbouring access optimisations. Targeting ExaSlang, ExaStencils \cite{exastencils} is a code generator to generate all optimised lower layers codes to automatically get optimal configurations and implementations.
The above tools cover a wide range of domains, designs, development platforms, languages and hardware targeted, optimisations and parallelisation strategies.
However, they lack the abstraction of the whole picture of a CFD problem. While stencils are highly optimised, time integration methods or ICs and BCs are not part of the language, leaving users to code the complete parallel workflow.

The second approach uses a top-level abstraction expressing the CFD problem in terms of actual differential equations. Those tools focus on a particular set of parallel numerical methods, leaving the implementation's details to lower-level libraries.
This is the case of FEniCS \cite{fenics}, a complete simulation infrastructure for many real-world problems.
FEniCS relies upon expressing PDEs at the mathematical level using Python and C++ interfaces.
The tool defines a language for the Finite Element Method and generates parallel codes that can be executed in parallel using MPI.
Although FEniCS is a compelling solution for many complex problems, it requires deep expertise in numerical methods to be used.
Similarly, Firedrake \cite{firedrake} uses the PyOP2 library for parallelising user high-level code.
OpenSBLI \cite{openSBLI} focus on the solution of the compressible Navier-Stokes equations. The tool uses symbolic Python to allow the specification of PDEs using Einstein notation which automatically discretises to generate OPS code.
Devito \cite{devito} uses similar symbolic Python mainly focusing on seismic inversion problems. Devito optimisations include common sub-expression elimination, vectorisation, blocking and multi-core and distributed parallelism.

While the first type of tools lacks part of the PDEs resolution process, the last usually rely on other frameworks to apply low-level optimisations and parallelisation strategies. This top-down dependence can worsen the use of advanced parallel paradigms such as the task model.
Saiph comprehends the combination of the above-targeted philosophies. It offers a CFD high-level syntax hiding numerical complexities while combining low-level automatic optimisations and parallelisation strategies leading to high-performance executions codes. As the above tools, Saiph design is based on layers separating concerns, however, the proposed execution model ensures the interaction between the high-level syntax layer and the low-level modules enabling to intertwine user code and low-level optimisations and parallelisation strategies.

\section{Conclusions}
In this work, we have enhanced Saiph, a high-level DSL for solving CFD problems using FDM, to generate optimised algorithmic patterns that leverage state-of-the-art optimisation techniques.
This work demonstrates how to enable and combine the most relevant optimisation techniques from a high-level specification code.
We extend Saiph to exploit high-performance algorithmic patterns found in FDM.
While maintaining abstraction and generality, we establish the procedure to generate codes exploiting single, multi-core, and cluster resources:
Saiph ensures high efficient sequential runs enabling compiler optimisations and vectorisation through automated compile-time evaluations, data-alignment and padding strategies.
Moreover, the DSL transparently provides scalable parallel codes preserving single-core optimisations;
using an appropriate blocking technique Saiph offers enough and well-balanced multi-core parallelism while enhancing data locality and preserving data alignment at each parallel chunk.
Tasking and fork-join parallelisation versions are available.
Saiph automatically distributes the mesh across several nodes but preserving the optimisations applied at the core and node level. 
Thus, following a bottom-up approach, we consciously combine such strategies from single-core to cluster level, preserving their effectiveness and obtaining high performance, competitive with hand-tuned codes.
We conclude that exploiting domain knowledge allows DSLs to assume and retrieve information from the high-level code.
Then, generic underlying implementations of optimisations and parallelisation strategies can be intertwined with the user code through the build process proposed.
In such a scenario, we automatically generated optimised codes using advanced parallelisation strategies and different parallel paradigms performing and scaling as much as manually optimised ones.
Further, due to the customisable nature of Saiph, it is possible to conduct a performance exploration under different parallelisation strategies.
Looking forward, we see the opportunity of transparently applying the best strategy for each input application.

\section*{Acknowledgment}
This research has received funding from the European Union's Horizon 2020/EuroHPC research and innovation programme under grant agreement N.955606 (DEEP-SEA), and is supported by the Spanish State Research Agency - Ministry of Science and Innovation (contract PID2019-107255GB), and by the Generalitat de Catalunya (2017-SGR-1414). This work is also supported by the Ministry of Economy of Spain through Severo Ochoa Center of Excellence Program (SEV-2015-0493).

\bibliographystyle{elsarticle-num}
\bibliography{bib/bibSaiph}

\end{document}